\documentclass[a4paper]{jpconf}
\usepackage{graphicx}
\begin{document}
\title{Testing the Pauli Exclusion Principle for Electrons}
\author{J.~Marton$^1$, S.~Bartalucci$^2$, S.~Bertolucci$^4$, C.~Berucci$^{1,2}$,
  M.~Bragadireanu$^{2,3}$, M.~Cargnelli$^1$, C. Curceanu (Petrascu)$^{2,3}$, S.~Di~Matteo$^5$, J.-P.~Egger$^6$,
  C.~Guaraldo$^2$, M.~Iliescu$^2$, T.~Ishiwatari$^1$, M.~Laubenstein$^7$,
E.~Milotti$^8$, D.~ Pietreanu$^2$, K.~Piscicchia$^{2,9}$, T.  Ponta$^{2,3}$, A.~ Romero~Vidal$^2$, A.~Scordo$^2$,
D.L.~Sirghi$^{2,3}$, F. Sirghi$^{2,3}$, L.~Sperandio$^2$, O.~Vazquez Doce$^2$, E.~Widmann$^1$ and J.~Zmeskal$^1$}

\address{$^1$ The Stefan Meyer Institute for Subatomic Physics, Boltzmanngasse 3, A-1090 Vienna, Austria}

\address{$^2$ INFN, Laboratori Nazionali di Frascati, CP 13,
 Via E. Fermi 40, I-00044, Frascati (Roma), Italy}

\address{$^3$ "Horia Hulubei" National Institute of Physics and
 Nuclear Engineering, \\ Str. Atomistilor no. 407, P.O. Box MG-6,
Bucharest - Magurele, Romania}

\address{$^4$ CERN, CH-1211, Geneva 23, Switzerland}

\address{$^5$ Institut de Physique UMR CNRS-UR1 6251, Universit\'e de Rennes1, F-35042 Rennes, France}

\address{$^6$ Institut de Physique, Universit\'e de Neuch\^atel, 1 rue A.-L. Breguet, CH-2000 Neuch\^atel, Switzerland}

\address{$^7$ Laboratori Nazionali del Gran Sasso, S.S. 17/bis, I-67010 Assergi (AQ), Italy}

\address{$^8$ Dipartimento di Fisica, Universit\`{a} di Trieste and INFN-- Sezione di Trieste, Via Valerio, 2, I-34127
Trieste, Italy}

\address{$^9$ CENTRO FERMI, Compendio del Viminale, Piazza del Viminale 1, I-00184 Roma, Italy}

\ead{johann.marton@oeaw.ac.at}

\begin{abstract}
One of the fundamental rules of nature and a pillar in the foundation of quantum theory
and thus of modern physics is represented by the Pauli Exclusion Principle.
We know that this principle is extremely well fulfilled due to many observations. Numerous experiments were performed to search for tiny violation of this rule in various systems. The experiment VIP at the Gran Sasso underground laboratory is searching for possible small violations of the Pauli Exclusion Principle for electrons leading to forbidden X-ray transitions in copper atoms. VIP is aiming at a test of the Pauli Exclusion Principle for electrons with high accuracy, down to the level of 10$^{-29}$ - 10$^{-30}$, thus improving the previous limit by 3-4 orders of magnitude.
The experimental method, results obtained so far and new developments within VIP2 (follow-up experiment at Gran Sasso, in preparation) to further increase the precision by 2 orders of magnitude will be presented.

\end{abstract}

\section{Introduction}
The Pauli Exclusion Principle (PEP) uncovered by Wolfgang Pauli in 1925 \cite{pauli25} is one of the corner stones of quantum physics and thus it is at the basis of the foundation of modern physics. It is connected with spin statistics dividing the world in fermions and bosons. Therefore, PEP is one of the most important rules in physics. Based on it is our understanding of nature and the consequences for the world of elementary particles up to compact objects (e.g. neutron stars) in the universe - but it is lacking a simple explanation as already stated by Pauli himself. Wolfgang Pauli showed later in 1940 the interconnection of the principle with the spin-statistics \cite{pauli40}. We know that the Pauli Principle is extremely well fulfilled but the the limit of validity - if any - is still an open question. Even a tiny violation of PEP would point to new physics which could show up at very high energies (at the Planck scale).
In the last decades several experimental tests of PEP validity for different systems have been performed \cite{bernabei,borexino,hilborn,nemo,nolte,tsipenyuk}.
A method to experimentally test the PEP was developed by Ramberg-Snow \cite{ramberg}. PEP is tested for {\it fresh} electrons, i.e. elementary particles having no interaction with the studied system thus circumventing the Messiah-Greenberg super-selection rule\cite{greenberg1964}. These {\it fresh} electrons are provided by a strong electric current which is flowing through a solid metal conductor. Pauli-forbidden transitions in this metal (e.g. K transitions from the 2p state to the 1s state already filled with 2 electrons) exhibit an significant energy shift in the transition energy which is resolvable by x-ray spectroscopy. A search for these x-ray transition events can be performed with high sensitivity but requires substantial background discrimination. It has to be noted that the {\it normal} K transitions are present due to background radiation.\\
In our previous VIP experiment at the underground laboratory LNGS (Laboratorio Nazionali di Gran Sasso) we used an improved Ramberg-Snow experimental setup exploiting charge coupled devices as x-ray detectors surrounding a solid cylindrical copper target. In this experiment we could deduce an upper limit for the Pauli exclusion principle violation in the order of 10$^{-29}$ \cite{Curceanu2012}.
A strongly improved VIP2 experiment will be performed in the underground laboratory LNGS in Gran Sasso taking again advantage of the excellent shielding against cosmic rays. A strongly improved compact setup with passive and active shielding will be used. Silicon drift detectors (SDDs) will serve as x-ray detectors providing a timing signal used in anti-coincidence with scintillators (i.e. veto counters) to suppress actively background events.\\
With the VIP2 experiment we want to improve the limit for PEP violation for electrons by 2 orders of magnitude reaching the range of 10$^{-31}$.

\section{Experimental Method}

The basic principle of the VIP experiment follows the method suggested by Greenberg and Mohapatra \cite{greenberg87} and is based on introducing {\it fresh} electrons into a copper strip. By applying an electrical current on the strip and searching for the Pauli-forbidden radiative transitions to the ground state that is already filled by two electrons. We look for 2p-1s transitions in copper which are easily detectable with x-ray detectors.
The energy of these non-Paulian transitions is shifted from the normal transition energy by about 300 eV (7.729 keV instead of 8.040 keV) due to the additional screening effect given by the two electrons in the 1s level \cite{sperandio}.
The measurements without the current in the copper strip provide the X-ray background, where no PEP violating transitions are expected. Those with the current provide the forbidden X-ray transitions when the {\it fresh} electrons could lead to Pauli-forbidden transitions. Comparing the two energy spectra, with and without the currents, the limit of the probability of the violation of PEP is extracted.

\subsection{VIP experiment}
The VIP experiment was installed in the underground laboratory of LNGS. This experiment used a copper cylinder as target and charge coupled devices (CCDs) as x-ray detectors. The array of CCD detectors were used for studies of kaonic atoms \cite{DEAR,ishiwatari}.
Details of the VIP experiment can be found in ref.\cite{vip,vipart}.
The measurement by the VIP experiment improved very significantly the limit previously obtained by Ramberg and Snow \cite{ramberg}, thanks to the following features:(a) use of CCD detectors instead of gaseous ones, having much better energy resolution (4-5 times better) and higher stability; (b) experimental setup located in the clean, low-background, environment of the underground LNGS Laboratory; and (c) collection of much higher statistics (longer DAQ periods, thanks to the stability of CCDs). We made full use of these features to obtain an improvement of several orders of magnitude on the previous limit by Ramberg and Snow (see tab.1).

\begin{table}
\caption{\label{VIPresults}Limits  of the Pauli violation probability for electrons from different experiments.}
\begin{center}
\begin{tabular}{llll}
\br
Experiment&Target&Upper limit of $\beta^2$/2 & ref.\\
\mr
Ramberg-Snow& Copper& 1.7x10$^{-26}$& \cite{ramberg}\\
S.R. Elliott et al.&Lead&1.5x10$^{-27}$ & \cite{elliott}\\
VIP(2006)&Copper&4.5x10$^{-28}$ & \cite{vipart}\\
VIP(2012)&Copper&4.7x10$^{-29}$ & \cite{Curceanu2012,laura}\\
VIP2(goal)&Copper&10$^{-31}$ & \\
\br
\end{tabular}
\end{center}
\end{table}

\subsection{VIP2 - a new high sensitivity experiment}

In order to achieve the signal/background increase which will allow a gain of two orders of magnitude for the probability of PEP violation for electrons, we are planning to build a new target, a new cryogenic system, use new detectors with timing capability and active veto system. As x-ray detectors we will use SDDs which were employed in der SIDDHARTA experiment on kaonic atoms at the DAFNE electron-positron collider of Laboratori Nazionali di Frascati. SDDs have an even better energy resolution than CCDs but additionally provide timing capability \cite{marton}which allow to use anti-coincidence operation with scintillators and therefore active shielding.
The VIP2 system will provide \\

\begin{itemize}
  \item signal increase with a more compact system with higher acceptance and higher current flow in the new copper strip target
  \item background reduction by decreasing the x-ray detector surface, more compact shielding (active veto system and passive), nitrogen filled box for radon radiation reduction
\end{itemize}

In the following table the numerical values for the improvements in VIP2 are given which will lead to an expected overall improvement of a factor higher than $\sim$120.

\begin{table}
\begin{center}
\caption{\label{factors}List of numerical values of the changes in VIP2 in comparison to the VIP features (given in brackets)}
\begin{tabular}{lll}
  \br
  Changes in VIP2 & value VIP2 (VIP) & expected gain\\
  \mr
  acceptance & 12\%  & 12 \\
  increase current & 100A (50A) & 2 \\
  reduced length & 3 cm (8.8 cm) & 1/3\\
  \br
  total linear factor && 8\\
  \br
  energy resolution & 170 eV (340 eV) & 4 \\
  reduced active area & 6 cm$^{2}$ (114 cm$^{2}$) & 20 \\
  better shielding and veto&  & 5-10 \\
  higher SDD efficiency &  & 1/2 \\
   \br
  background reduction &  & 200-400 \\
   \br
  overall improvement &  & $>$ 120\\
  \br
\end{tabular}
\end{center}
\end{table}

\begin{figure}
\begin{center}
\includegraphics[width=4in]{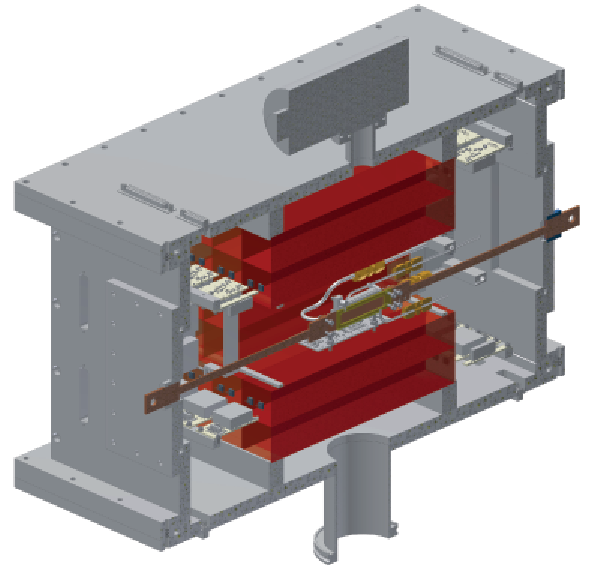}
\end{center}
\caption{\label{VIP2arrangement}An artist view of the VIP2 experimental setup. In the middle the copper conductor and the x-ray detectors are installed. Plastic scintillators with solid state photodetector readout acting as active shielding (see fig.\ref{scintillator}) are surrounding this inner part.}
\end{figure}

\begin{figure}
\begin{center}
\includegraphics[width=4in]{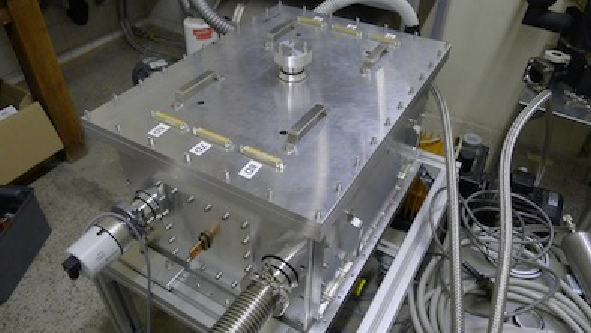}
\end{center}
\caption{\label{VIP2photo}Photo of the VIP2 box assembled for first tests in the laboratory.}
\end{figure}

Figures \ref{VIP2arrangement} and \ref{VIP2photo} show the main elements for the proposed setup in the VIP2 experiment. The copper strip target is 30 mm long, 10 mm wide and is about 40$\mu$m  thick, and is installed in the center of the setup. The copper strip is cooled at $\sim$90K by the use of an external cryogenic system using liquid argon as the cooling medium. The current connection lines made of copper wires with a cross-section area of 1.5 cm$^{2}$, allow a current flow of (at least) 100 A. The current lines exhibit a temperature gradient from inside the vacuum chamber to the outside connectors of about 180 K.

Monte Carlo simulations were performed to study the effect of the active shielding in various configurations of the setup and models of the background radiation. The background profiles measured at LNGS \cite{bucci} were used as input parameters in the simulations.

\begin{figure}
\begin{center}
\includegraphics[width=3in]{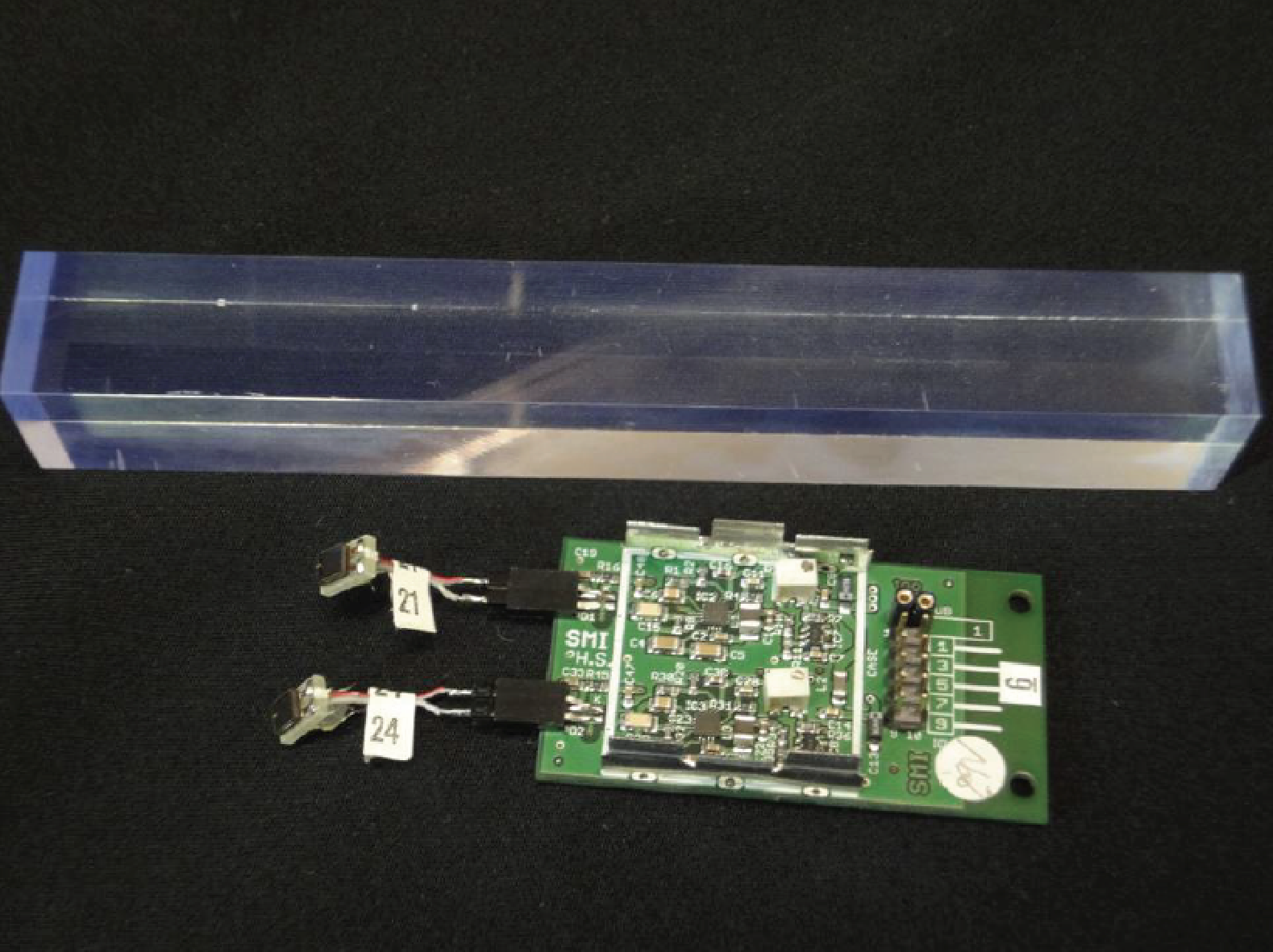}
\end{center}
\caption{\label{scintillator}Component from the active veto system of VIP2: Plastic scintillator bar with SiPM readout electronic board.}
\end{figure}

\begin{figure}
\begin{center}
\includegraphics[width=5in]{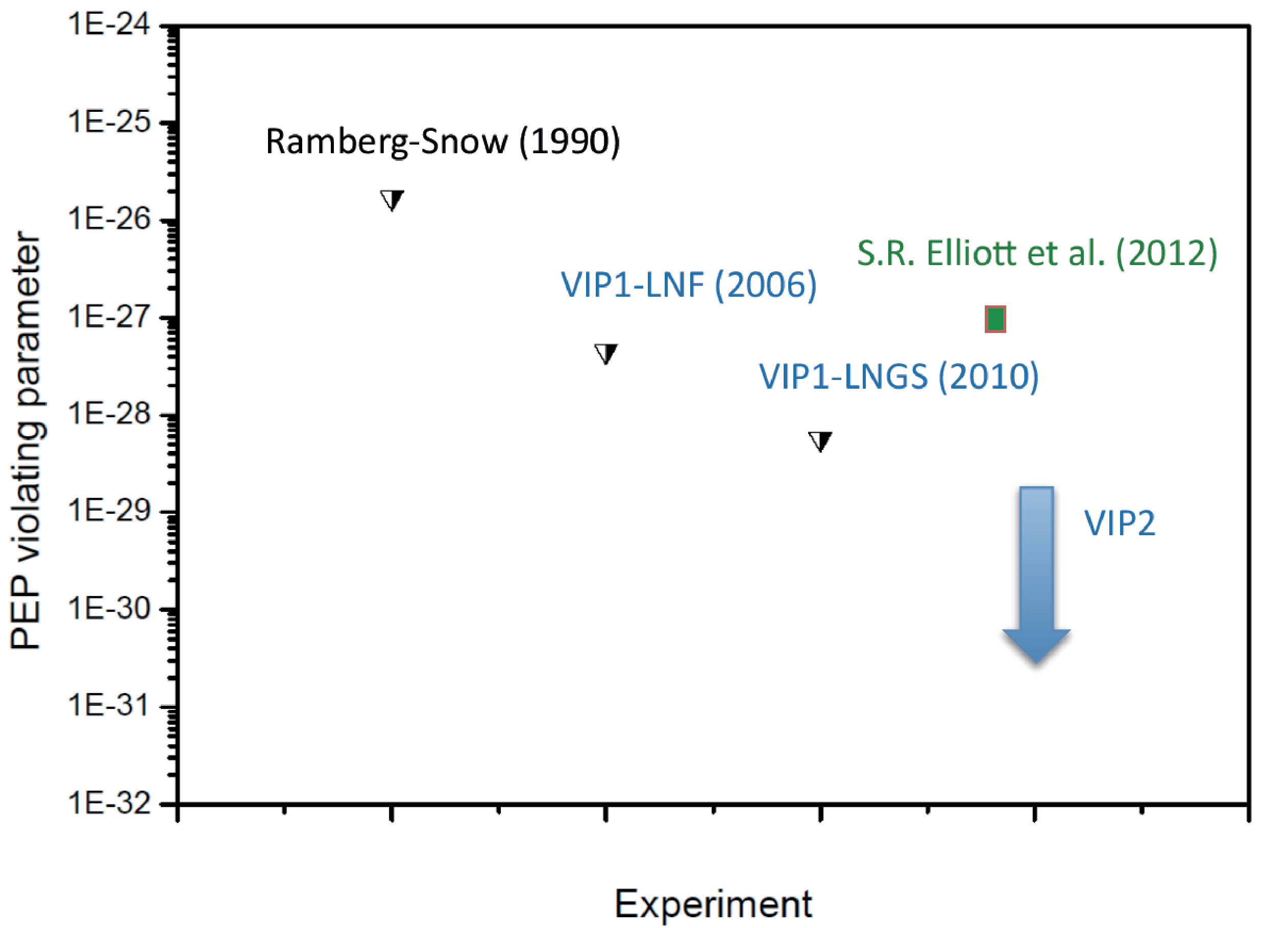}
\end{center}
\caption{\label{Results-beta}Results of PEP violation experiments for electrons.}
\end{figure}

 As the veto counters (see fig.\ref{scintillator}), 2 pieces of 10 cm thick plastic scintillators are found to be optimal. We also simulated a setup with BGO inorganic crystals which yields even better background suppression, but due to the high costs of this solution, we decided to propose plastic scintillators, e.g. Bicron 412.
To summarize we are confident that the goal of VIP2 to reach an upper limit for possible PEP violation in the order of 10$^{-31}$ can be reached with the improved setup
(see fig.\ref{Results-beta}).

\section{Summary and Outlook}

In the light of the importance of the Pauli Exclusion Principle high precision tests of this rule of nature are well justified. Especially interesting are cases in which new
fermions are used as test particles - in the case of VIP2 fresh electrons introduced by an electric current. In VIP2 the background will be reduced by a factor of 200-400 resulting in a gain of $\sim$ 15-20 in the expected upper limit for $\beta^{2}$/2. Together with the gain in signal the total gain improvement will excceed 2 orders of magnitude. \\
 Therefore, in the VIP2 experiment the limit of the PEP violation can be reached to 10$^{-31}$, which is two orders of magnitude better than the previous limit of $\sim$4x10$^{-29}$. This limit is anticipated to be obtained taking into account a data taking period similar to that of VIP1, i.e. about 3 years).
 This incredible precision might help to prove or disprove tiny violations related to {\it new physics} like e.g. superstring motivated effects \cite{Jackson}.

\subsection*{Acknowledgement}

The very important assistance of the INFN-LNGS laboratory staff during all phases of preparation, installation and data
taking as well as the support from the HadronPhysics FP6 (506078), HadronPhysics2 FP7 (227431), HadronPhysics3 (283286) projects and the EU COST 1006 Action is gratefully
acknowledged. Especially we thank the Austrian Science Foundation (FWF) which supports the VIP2 project with the grant P25529-N20.

\end{document}